\documentclass[11pt,a4paper]{article}
\usepackage{jheppub}
\title{Canonical Analysis of Inhomogeneous Dark Energy Model and Theory of Limiting
Curvature }

\author{Josef Kluso\v{n}}

\affiliation{Department of Theoretical Physics and Astrophysics,
Faculty of Science,\\ Masaryk University, Kotl\'a\v{r}sk\'a 2, 611 37,
Brno, Czech Republic}

\emailAdd{klu@physics.muni.cz}

\abstract{This paper is devoted to the canonical analysis of 
inhomogeneous Dark Energy Model  and the model of limiting curvature that were proposed recently  by Chamseddine and V. Mukhanov. 
We argue these models are well defined and have similar properties 
as a   system  consisting from   general gravity action  and action 
for  incoherent dust.}


\def\tmH{\tilde{\mH}}

\def\mH{\mathcal{H}}

\def\bx{\mathbf{x}}
\def\by{\mathbf{y}}

\newcommand{\mK}{\mathcal{K}}

\newcommand{\tk}{\tilde{k}}

\newcommand{\mG}{\mathcal{G}}

\newcommand{\bT}{\mathbf{T}}

\newcommand{\mL}{\mathcal{L}}

\def\pb #1{\left\{#1\right\}}

\newcommand{\phys}{\mathrm{phys}}

\begin{document}
\maketitle
\flushbottom

\section{Introduction}
Few years ago a  new interesting model of mimetic dark matter was suggested
in \cite{Chamseddine:2013kea} and was further elaborated in
\cite{Barvinsky:2013mea,Chamseddine:2014vna}
\footnote{For review and extensive list of references, see   \cite{Sebastiani:2016ras}.}. In mimetic gravity it is possible
to describe the dark components of the Universe as a purely geometrical effect, without the need of introducing additional matter fields. This
description can be achieved using very simple but remarkable idea.
 The physical metric $g_{\mu\nu}^{\phys}$ is
considered to be a function of a scalar field $\phi$ and a fundamental
metric $g_{\mu\nu}$, where the physical metric is defined as\footnote{We
follow the convention used in \cite{Barvinsky:2013mea} and we also
consider the space-time metric of the signature $(-,+,+,+)$.}
\begin{equation}\label{g^phys}
g_{\mu\nu}^{\phys}=\left(-g^{\alpha\beta}\partial_\alpha\phi
\partial_\beta\phi\right) g_{\mu\nu} \  .
\end{equation}
The physical metric $g_{\mu\nu}^{\phys}$ is invariant with respect to
the Weyl transformation of the metric $g_{\mu\nu}$,
\begin{equation}\label{defg'}
g'_{\mu\nu}(x)=\Omega^2(x) g_{\mu\nu}(x) \  .
\end{equation}
Then it was shown in \cite{Chamseddine:2013kea} and in
\cite{Barvinsky:2013mea} that the ordinary Einstein-Hilbert action
constructed using the physical metric $g_{\mu\nu}^{\phys}$ possesses
many interesting properties since this model can be interpreted as a 
conformal extension of Einstein's general theory of relativity. The
local Weyl invariance is ensured by introducing an extra degree of
freedom that as was shown in \cite{Chamseddine:2013kea} has the form of
pressureless perfect fluid that, according to
\cite{Chamseddine:2013kea}, can mimic the behavior of a real cold dark
matter. 

This mimetic dark matter proposal \cite{Chamseddine:2013kea}
was recently generalized to so called "inhomogeneous dark energy model"
in \cite{Chamseddine:2016uyr}. In this paper the original mimetic 
model was extended in oder to describe arbitrary inhomogeneous 
dark energy in any scale and that can contribute to the gravitational instability at late time. This theory is described by the action 
\footnote{The inhomogeneous extension of mimetic gravity is based
on Einstein frame version of this theory where the manifest Weyl invariance  (\ref{defg'})
is lost. For more details, see for example \cite{Chaichian:2014qba}. }
\begin{equation}\label{SgfEgen}
S=\frac{1}{2}\int d^4x\sqrt{-g}
\left[ R(g_{\mu\nu}) -\lambda\left( 1+g^{\mu\nu}
\nabla_\mu\phi\nabla_\nu\phi \right)-\lambda_a g^{\mu\nu}
\partial_\mu\phi^a\partial_\nu\phi-V(\phi^a) \right] \ ,
\end{equation}
where $a=1,\dots,D$ where $D$ counts the number of the scalar fields $\phi^a$. 
Generally we can have $D$ arbitrary but the simplest possibility is 
to take $a=1$ while the most convenient case is to consider $D=3$ 
\cite{Chamseddine:2016uyr} since then we can   identify the scalars with 
the synchronous coordinates. The variation of the action with respect to Lagrange multipliers $\lambda$ and $\lambda_a$ gives following equations of motion 
\begin{equation}
\frac{\partial \phi}{\partial x^\mu}\frac{\partial \phi}
{\partial x^\nu}g^{\mu\nu}=-1 \ , \quad 
\frac{\partial \phi^a}{\partial x^\mu}\frac{\partial \phi}
{\partial x^\nu}g^{\mu\nu}=0 \ .   
\end{equation}
 It was shown in \cite{Chamseddine:2016uyr}
that at synchronous coordinate system where $g_{00}=-1 \ , 
g_{0i}=0 \ , i=1,2,3$ so that 
\begin{equation}
ds^2=-dt^2+h_{ij}dx^i dx^j 
\end{equation}
the general solution of these equations in this coordinate
system are
\begin{equation}
\phi=t \ , \phi^a=\phi^a(x^i) \  
\end{equation}
so that the fields $\phi^a$ are time independent 
functions and the potential can be time independent function 
of the spatial coordinates and it leads to cosmological-like
constant in the Einstein equations. It was also argued
here that the inhomogeneities
 in the distribution of dark 
energy can have an impact on the power spectrum at large
scales and the formation of structure of the universe. 

All these facts makes the proposal  \cite{Chamseddine:2016uyr}
very interesting and  deserves further study. Certainly 
it would be very useful to find Hamiltonian formulation of this theory. In fact,  canonical analysis of the original mimetic dark matter model was very carefully performed in 
\cite{Chaichian:2014qba}
\footnote{See also \cite{Malaeb:2014vua}.} where we identified 
all constraints and determined number of physical degrees of freedom.
We also shown that by  solving the second class constraints
the resulting Hamiltonian for the scalar field is linear in the momentum 
conjugate to the scalar field. The presence of the linear momentum
signals instability of the theory since the Hamiltonian is not bounded
from below for
certain type of initial configurations and consequently it can become
unstable.

There is a natural question how the situation changes when we consider 
the model presented in \cite{Chamseddine:2016uyr} and this is  one of the goals
of the present paper. It turns out that the Hamiltonian analysis is rather non-trivial
and depends on the number of  additional scalar fields $\phi^a$. In case when 
$a\neq 1$ we find new primary constraints that have to be taken into account. 
Then we also derive corresponding secondary constraints. Despite of the
complexity of the Hamiltonian analysis we find that after solving second class
constraints  the Hamiltonian constraint is linear in momentum conjugate
to the scalar field $\phi$.  On the other hand we show that this contribution can be rewritten to the form corresponding to the Hamiltonian 
constraint for the dust \cite{Brown:1994py} which is well defined
system with quadratic Hamiltonian bounded from below. This analysis 
shows that the mimetic model and model studied here could be stable.

The final part of our paper is devoted to the Hamiltonian analysis of the model
that was introduced very recently in \cite{Chamseddine:2016uef,Chamseddine:2016ktu} which is very interesting proposal how to resolve singularities in general relativity. It is based on the idea of  modification of classical general  relativity at high curvatures by incorporating 
limiting curvature. If this limiting curvature is few order
of magnitude below the Planckian value  we can ignore 
quantum gravity effects. This remarkable proposal is again based on 
the existence of the scalar field that obeys the equation
\begin{equation}\label{mimconst}
g^{\mu\nu}\partial_\mu\phi\partial_\nu\phi=-1 \ . 
\end{equation}
It was also presumed that the theory is invariant under 
shift symmetry $\phi\rightarrow \phi+\mathrm{const}$ 
which implies that the potential $V(\phi)$ is absent. 
On the other hand a  new additional term $f(\Box \phi)$ was
added to the action, where the function $f$ was suggested to 
have the form 
\begin{equation}\label{fBox}
f(\Box \phi)=1-\sqrt{1-\frac{(\Box \phi)^2}{\epsilon_m}}+
\dots \ . 
\end{equation}
It was shown in \cite{Chamseddine:2016uef,Chamseddine:2016ktu} that
this theory resolves singularities in Friedmann and Kasner universes. In other words the contracting universes bounce at the limiting 
curvature and all curvature invariants are regular and bounded
by the values characterized by $\epsilon_m$. Then it was shown in 
\cite{Chamseddine:2016ktu} that the physical singularity 
of the Schwarzschild black hole can be removed as well. 

Since this proposal is based on an extension of mimetic action 
with a new term  that 
depends on $\Box\phi$ we should study whether these  higher
derivatives do not imply  instability of this theory.
 In order to answer this question  we perform Hamiltonian analysis of this theory. 
By introducing two auxiliary fields we rewrite the action to the 
form that contains the first order derivatives only
and hence it is suitable for the Hamiltonian analysis. We determine corresponding 
Hamiltonian and we also find structure of the constraints. Surprisingly we find that the presence of the higher derivative term does not 
lead to the existence of additional  degree of freedom. This 
fact is a consequence of the presence  of the  constraint 
(\ref{mimconst})  in the action which implies that this theory is
degenerate while Ostrogradsky's theorem is strictly speaking valid for non-degenerate theories only
\footnote{For recent review of Ostrogradsky's theorem, see 
\cite{Woodard:2015zca}.}.

In conclusion,  we mean that  the proposal suggested in 
\cite{Chamseddine:2016uef,Chamseddine:2016ktu} is very remarkable 
and deserves further study. In particular, these theories could 
be useful for "deparameterising of the theory of gravity" \cite{Thiemann:2006up,Brown:1994py}. Briefly, this idea is based on a presumption that the Hamiltonian constraint can be written in the form $\mH(\bx)=\pi(\bx)+\mK(\bx)$, where $\pi(\bx)$ is the momentum conjugate to the scalar field $\phi(\bx)$ and where $\mK$ is positive function on phase space which depends neither on $\phi$ or $\pi$.  
Then it is possible to construct physical observable and the function 
$\mathbf{K}=\int d^3\bx \mK(\bx)$ is the natural physical Hamiltonian
that generates the time evolution of the observables, see 
\cite{Thiemann:2006up}. It would be very interesting to see whether 
 models presented in \cite{Chamseddine:2016uyr,Chamseddine:2016uef,Chamseddine:2016ktu}
could allow such a construction. We currently analyze this problem. 

The structure of this paper is as follows. In the next section (\ref{second}) we perform Hamiltonian analysis of Inhomogeneous Dark Energy model as was formulated in  \cite{Chamseddine:2016uyr}.
Then in section (\ref{third}) we consider simpler formulation of this model when the number of additional scalar fields is equal to one. In section (\ref{fourth}) we perform Hamiltonian analysis of the model 
\cite{Chamseddine:2016uef,Chamseddine:2016ktu}. Finally in conclusion 
(\ref{fifth}) we outline our results and suggest possible extension 
of this work. 

\section{Hamiltonian Analysis of Inhomogeneous Dar Energy model}\label{second}
In this section we perform Hamiltonian analysis of the action 
(\ref{SgfEgen})
in the full generality when we will presume that  $a=1,\dots,D$.
In order to find its Hamiltonian form we  use  the following $3+1$ decomposition of the metric $g_{\mu\nu}$
\cite{Arnowitt:1962hi,Gourgoulhon:2007ue}
\begin{eqnarray}\label{hgdef}
& &g_{00}=-N^2+N_i h^{ij}N_j , \quad g_{0i}=N_i , \quad
g_{ij}=h_{ij} ,\nonumber \\
& &g^{00}=-\frac{1}{N^2} , \quad g^{0i}=\frac{N^i}{N^2} ,
\quad g^{ij}=h^{ij}-\frac{N^i N^j}{N^2} ,
\end{eqnarray}
where we have defined $h^{ij}$ as the inverse to the induced metric
$h_{ij}$ on the Cauchy surface $\Sigma_t$ at each time $t$,
\begin{equation}
h_{ik}h^{kj}=\delta_i^{ \ j},
\end{equation}
and we denote $N^i=h^{ij}N_j$. The four dimensional scalar curvature in
$3+1$ formalism has the form
\begin{equation}
R(g_{\mu\nu})=
K_{ij}\mG^{ijkl}K_{kl}+R+\frac{2}{\sqrt{-g}}\partial_\mu(\sqrt{-g}
n^\mu K) -\frac{2}{\sqrt{h}N}\partial_i(\sqrt{h}h^{ij}\partial_j N) ,
\end{equation}
where the extrinsic curvature of the spatial hypersurface $\Sigma_t$ at
time $t$ is defined as
\begin{equation}
K_{ij}=\frac{1}{2N}\left(\frac{\partial h_{ij}}{\partial t}- D_i
N_j-D_j N_i\right) ,
\end{equation}
with $D_i$ being the covariant derivative determined by the metric
$h_{ij}$, and where the de Witt metric is defined as
\begin{equation}
\mG^{ijkl}=\frac{1}{2}(h^{ik}h^{jl}+h^{il}h^{jk}) -h^{ij}h^{kl}
\end{equation}
with inverse
\begin{equation}
\mG_{ijkl}=\frac{1}{2}(h_{ik}h_{jl}+h_{il}h_{jk}) -\frac{1}{2}
h_{ij}h_{kl} \
\end{equation}
that obeys the relation
\begin{equation}
\mG_{ijkl}\mG^{klmn}=\frac{1}{2}(\delta_i^m\delta_j^n+\delta_i^n
\delta_j^m) .
\end{equation}
Further, $n^\mu$ is the future-pointing unit normal vector to the
hypersurface $\Sigma_t$, which is written in terms of the ADM variables
as
\begin{equation}
n^0=\sqrt{-g^{00}}=\frac{1}{N} ,\quad
n^i=-\frac{g^{0i}}{\sqrt{-g^{00}}}=-\frac{N^i}{N} .
\end{equation}
Note that we are not interested in the boundary terms so that the terms proportional 
to total derivatives are not important for us and can be ignored. 

Now inserting this $3+1$ decomposition into the action
(\ref{SgfEgen})
we obtain
\begin{eqnarray}\label{actHam}
S&=&\frac{1}{2} \int dt d^3\bx\sqrt{h}N
[ K_{ij}\mG^{ijkl}K_{kl} +R+\lambda(-1+\nabla_n\phi\nabla_n\phi-h^{ij}
\partial_i\phi\partial_j\phi)+\nonumber \\
&+&\lambda_a \nabla_n\phi^a\nabla_n\phi-\lambda_a h^{ij}\partial_i\phi^a\partial_j\phi-V(\phi^a)] \ , 
\nonumber \\
\end{eqnarray}
where
\begin{equation}
\nabla_n\phi=\frac{1}{N}(\partial_t\phi-N^i\partial_i\phi) \  ,
\end{equation}
and where we ignored boundary terms.
Before we proceed further we should also stress one important point. Since $V(\phi^a)$ is a scalar function of $\phi^a$ the only possibility how to construct scalar from 
the vectors $\phi^a$ is to perform contractions of these two vectors. In order to do this we have to introduce 
general metric $\omega_{ab}$ on the space spanned by $\phi^a$. We will presume that $\omega_{ab}$ is constant with inverse $\omega^{ab}$. Of course, the simplest possibility is $\omega_{ab}=\delta_{ab}$ but we will keep 
$\omega_{ab}$ general. Then $\lambda^a=\omega^{ab}\lambda_b$.

Now we can easily derive
the momenta conjugate to $h_{ij},\Phi,\lambda,\lambda_a$ and $\phi,\phi^a$ from the
action (\ref{actHam}) as
\begin{eqnarray}
\pi^{ij}&=&\frac{\delta L}{\delta \partial_t h^{ij}}=\frac{1}{2}\sqrt{h}\mG^{ijkl}K_{kl} , \quad  \pi_N=\frac{\delta L}{\delta \partial_t N}\approx 0 \ , \quad \pi_i=\frac{\delta L}{\delta \partial_t N^i}\approx 0 \ ,  \nonumber \\
p_\phi&=&\frac{\delta L}{\delta\partial_t \phi}=\lambda \sqrt{h}\nabla_n\phi
+\frac{1}{2}\sqrt{h}\lambda_a \nabla_n\phi^a \ ,  \nonumber \\
  k^a&=&\frac{\delta L}{\delta \partial_t
\lambda_a}\approx 0 \ , \quad k=\frac{\delta L}{\delta \partial_t \lambda}\approx 0 \ ,   \nonumber \\
p_a&=&\frac{\delta L}{\delta \partial_t \phi^a}=\frac{1}{2}\sqrt{h}
\lambda_a \nabla_n\phi \  . \nonumber \\
\end{eqnarray}
Very interesting is the expression for $p_a$ since it implies 
\begin{equation}
\frac{\lambda_b \lambda^a}{\lambda^c\lambda_c}p_a=p_b
\end{equation}
that leads to the following set of $D-1$ primary constraints 
\begin{equation}
\Sigma_a=P_a^{ \ b}p_b\approx 0 \ , \quad  \lambda^a \Sigma_a=0 \  , 
\end{equation}
where $P_b^{ \ a}=\delta_b^a+\frac{\lambda_b\lambda^a}{\lambda^c \lambda_c}$ is the projector to the space orthogonal to one dimensional
space spanned by $\lambda_a$ since
\begin{equation}
P_b^{ \ a}\lambda_a=0 \ . 
\end{equation}
Now it is easy to determine corresponding bare
Hamiltonian 
\begin{eqnarray}
H&=&\int d^3\bx (\pi^{ij}\partial_t h_{ij}+p_\phi \partial_t\phi+p_a\partial_t\phi^a-\mL)=
\nonumber \\
&=&\int d^3\bx (N\mH_T+N^i\mH_i) \ ,  \nonumber \\
\end{eqnarray}
where
\begin{eqnarray}\label{mHT1}
\mH_T&=&\frac{2}{\sqrt{h}}\pi^{ij}\mG_{ijkl}\pi^{kl}-\sqrt{h}R+\frac{2}{\sqrt{h}
\lambda_a \lambda^a}p_\phi (\lambda^c p_c)-\frac{2\lambda}{\sqrt{h}(\lambda^a \lambda_a)^2
}(\lambda^b p_b)^2
\nonumber \\
&+&\frac{1}{2}\lambda\sqrt{h}(1+h^{ij}\partial_i\phi
\partial_j\phi)+\frac{1}{2}
\sqrt{h}\lambda_a h^{ij}\partial_i\phi^a\partial_j\phi+
\frac{1}{2}\sqrt{h}V(\phi^a) \ , \nonumber \\
\mH_i&=&-2h_{ik}D_j\pi^{kj}+p_\phi\partial_i\phi+p_a\partial_i\phi^a \ .
\nonumber \\
\end{eqnarray}
Before we proceed further we should stress one important point which 
is related to the fact that $k^a$ do not Poisson commute with projector $P_a^{ \ b}$. Explicitly we find following Poisson brackets 
\begin{eqnarray}
\pb{k^a(\bx),\Sigma_b(\by)}=\frac{\lambda^c p_c}{\lambda^d \lambda_d }
P^a_{ \ b}\delta(\bx-\by)  \ , \nonumber \\
\end{eqnarray}
that however also implies that 
\begin{equation}
\pb{\lambda_a k^a(\bx),\Sigma_b(\by)}=0 \ . 
\end{equation}
We see that it is natural to split $k^a$ into $\tk^a
\equiv k^a-\frac{\lambda^b k_b}
{\lambda^d \lambda_d}\lambda^a=P^a_{ \ b} k^b$
that is orthogonal to $\lambda_a$ and their complement which is projection 
of $k^a$ along $\lambda_a$ defined as 
\begin{equation}
\psi\equiv \lambda^a k_a \ . 
\end{equation}
 Note that $\tk^a$ has $D-1$ independent
components since it obeys $\tk^a \lambda_a=0$ by definition. It is important
that there are non-zero Poisson brackets between $\tk^a$ and $\Sigma_b$ equal to 
\begin{eqnarray}
\pb{\tk^a(\bx),\Sigma_b(\by)}
=\frac{\lambda^d p_d}{\lambda^e
	\lambda_e }P^a_{ \ b}\delta(\bx-\by) \ .  \nonumber \\
\end{eqnarray}
In other words $\tk^a$ and $\Sigma_a$ are sets of 
 $2(D-1)-$second class constraints. Then we introduce extended form of 
 Hamiltonian with  all  primary constraints included 
\begin{equation}\label{HT1}
H_T=\int d^3\bx (N\mH_T+N^i\tmH_i+v^\psi \psi+ v^k k+v^N\pi_N+
v^i\pi_i+w^a \Sigma_a+v_a \tk^a) \ , 
\end{equation}
where $v^\psi,v^k,v^N,v^i$ are unspecified Lagrange multipliers corresponding to the primary constrains. On the other hand $w^a, v_a$
belong to the space orthogonal to subspace generated by $\lambda_a$. 
Further, it is convenient to 
 extended $\mH_i$ in the following way
\begin{equation}
\tmH_i=\mH_i+k \partial_i\lambda+k^a\partial_i\lambda_a \ . 
\end{equation}
As the first step we analyze the requirement of preservation of the primary 
constraints during the time evolution of the system. In case of 
 the constraints $\tk^a$ and $\Sigma_a$ we obtain  
\begin{eqnarray}\label{reqtka}
\partial_t \tk^a&=&\pb{\tk^a,H_T}=\int d^3\bx \left(N\pb{\tk^a,\mH_T(\bx)}+
w^b(\bx)\pb{\tk^a,\Sigma_b(\bx)}\right)=0 \ , \nonumber \\
\partial_t \Sigma_a&=&\pb{\Sigma_a,H_T}=\int d^3\bx \left(N\pb{\Sigma_a,\mH_T(\bx)}+
v_b(\bx)\pb{\Sigma_a,\tk^b(\bx)}\right)=0 \ . \nonumber \\
\end{eqnarray}
Since $w^b$ belong to the subspace transverse to $\lambda_a$ we find that $w^b(\by) \pb{\tk^a(\bx),\Sigma_b(\by)}=\frac{\lambda^d p_d}
{\lambda^e\lambda_e}P_b^aw^b\delta(\bx-\by)=
\frac{\lambda^d p_d}{\lambda^a\lambda_e}w^a\delta(\bx-\by)$ and 
hence the first equation in (\ref{reqtka}) has the solution 
\begin{eqnarray}
w^a=\frac{1}{2}\frac{\lambda^e\lambda_e}{\lambda^d p_d}NP^a_{ \ b} \partial_j \phi^b h^{ij}\sqrt{h}
\partial_j\phi \  \nonumber \\
\end{eqnarray}
while the second one implies 
\begin{eqnarray}
v_a=\frac{1}{2}\frac{\lambda^e\lambda_e}{\lambda^d p_d}
P_a^{ \ b} \partial_j\lambda_b N\sqrt{h}h^{ij}\partial_j\phi \ . 
\nonumber \\
\end{eqnarray}
Since $\tk^a,\Sigma_a$ have
vanishing Poisson brackets with remaining  primary constraints they effectively decouple. 

Now we proceed to the analysis of the preservation of the primary
constraints $\pi_N\approx 0 , \pi_i\approx 0, k, \sigma\approx 0.$  As usual the requirement of the preservation of
the constraints $\pi_N,\pi_i$ implies the secondary constraints
\begin{equation}\label{mH=0}
\mH_T\approx 0 \ ,\quad \tmH_i\approx 0 \ .
\end{equation}
For further analysis we introduce the smeared form of these
constraints
\begin{equation}
\bT_T(N)=\int d^3\bx N\mH_T ,\quad \bT_S(N^i)= \int d^3\bx
N^i\tmH_i \ .
\end{equation}
The requirement of the preservation of the constraint $k\approx 0$ implies
\begin{equation}
\partial_t k=\pb{k,H}=N\left(\frac{2}{\sqrt{h}(\lambda^a \lambda_a)^2}(\lambda^b p_b)^2
-\frac{1}{2}\sqrt{h}(1+h^{ij}\partial_i\phi\partial_j\phi)\right)\equiv N\Omega\approx 0  \ . 
\end{equation}
Let us now proceed to the analysis of  time evolution of the constraint 
$\psi$. Since 
 $\psi$ has zero Poisson bracket with $
\Sigma_a$  the requirement of its preservation during  time evolution of the system implies new constraint. Explicitly, we find  
\begin{eqnarray}
\partial_t\psi=\pb{\psi,H_T}=
N\Sigma \approx 0 \ , \nonumber \\
\end{eqnarray}
where 
\begin{equation}
\Sigma=
\frac{2p_\phi (\lambda^a p_a)}{\sqrt{h}(\lambda^a \lambda_a)}-
\frac{4 \lambda (\lambda^b p_b)^2}{\sqrt{h}(\lambda^a\lambda_a)^2}
-\frac{1}{2}\sqrt{h}h^{ij}\lambda_a\partial_i\phi^a \partial_j\phi\approx 0 \ . 
\end{equation}
In summary we  have following set  of the second class
constraints $\Psi_A=(\tilde{k}^a,\Sigma_a,k,\psi,\Omega,\Sigma)$.
Now the matrix of  Poisson brackets between these second class constraints has schematic form
\begin{equation}
\triangle_{AB}\equiv \pb{\Psi_A,\Psi_B}=\left(\begin{array}{cccccc} 
0 & * & 0 & 0 & 0 & * \\
* & 0 & 0 & 0 & 0 & * \\
0 & 0 & 0 & 0 & * & * \\
0 & 0 & 0 & 0 & * & * \\
0 & 0 & * & * & * & * \\
* & * & * & * & * & * \\
\end{array}\right) \ , 
\end{equation}
where $*$ means non-zero elements. Then it is easy to see that
the inverse matrix has  schematic form 
\begin{equation}\label{traingleinvs}
\triangle^{AB}
=\left(\begin{array}{cccccc} 
* & * & * & * & 0 & 0 \\
* & * & * & * & 0 & 0 \\
* & * & * & * & * & * \\
* & * & * & * & * & * \\
0 & 0 & * & * & 0 & 0 \\
0 & 0 & * & * & 0 & 0 \\
\end{array}\right) \ . 
\end{equation}
Finally we determine Poisson brackets between $\mH_T$ and $\mH_i$.
We use their smeared form and we find
\begin{eqnarray}
\pb{\bT_T(N),\bT_T(M)}
&=&\bT_S((N\partial_iM-M\partial_iN)h^{ij})+
\int d^3\bx (N\partial_iM-M\partial_iN)h^{ij}\partial_j\phi^a \Sigma_a \ , 
\nonumber \\
\pb{\bT_S(N^i),\bT_T(M)}&=&\bT_T(N^i\partial_iM) \ , \nonumber \\
\pb{\bT_S(N^i),\bT_S(M^j)}&=&\bT_S(N^j\partial_j M^i-M^j\partial_j N^i) \ .  
\nonumber \\
\end{eqnarray}
Since all second class constraints are invariant under spatial diffeomorphism 
we find that they have weakly vanishing Poisson brackets with $\bT_S(N^i)$. On the other hand it is easy to see that there are non-zero Poisson brackets between $\mH_T$ and some of the constraints $\Psi_A$. Then it is convenient to introduce following constraint 
\begin{equation}\label{tmHT}
\tmH_T=\mH_T-\pb{\mH_T,\Psi_A}\triangle^{AB}\Psi_B \ , 
\end{equation}
where the sumation over $A$ includes integration over space coordinates. From 
(\ref{tmHT}) we easily find that  $\pb{\tmH_T,\Psi_B}=0$. This relation ensures that $\tmH_T$ is the first class constraint. In summary, we have four first class constraints $\tmH_T,\tmH_i$ as it is expected for the theory 
invariant under full diffeomorphism. Note also that $\tmH_T$ coincides with 
$\mH_T$ when all  second class constraints strongly vanish.  

Finally we determine the structure of the Dirac brackets. 
Let us denote the Poisson bracket between all canonical variables  
and the vector of the second class constraints in the schematic form 
\begin{eqnarray}
& &\pb{\phi^a,\Psi^T}=(0,*,0,0,*,*)
\ , 
\pb{\lambda_a,\Psi^T}=(*,0,0,0,0,0) \ , \nonumber \\
& &\pb{h_{ij},\Psi^ T}=(0,0,0,0,0,0) \ , \quad 
 \pb{\pi^{ij},\Psi^T}=(0,0,0,0,*,*) \ . \nonumber \\
& &\pb{p_a,\Psi^T}=(0,0,0,0,*,*) \ . \nonumber \\
\end{eqnarray}
Using this expression and (\ref{traingleinvs}) we easily find 
\begin{eqnarray}
& &\pb{\phi^a,p_b}_D=\pb{\phi^a,p_b}=\delta^a_b \ , 
\nonumber \\
& &\pb{\phi^a(\bx),\lambda_b(\by)}=\triangle^a_b(\bx,\by)  \ , \nonumber \\
& &\pb{\phi^a,\pi^{ij}}_D=\pb{\phi^a,h_{ij}}_D=0 \ , \nonumber \\
& &\pb{h_{ij},\pi^{kl}}_D=\pb{h_{ij},\pi^{kl}} \ , 
\pb{\pi^{ij},\pi^{kl}}=0 \ , \nonumber \\
& & \pb{\pi^{ij},\lambda_a}_D=0 \ . \nonumber \\
\end{eqnarray}
where $\triangle^a_b$ is a non-trivial matrix whose explicit form is not 
important for us. 
We see that there is non-trivial phase space structure between $\phi^a$ and 
$\lambda_b$ so that  $\lambda_b$ can be effectively considered as 
the variable conjugate to $\phi^a$. In fact, this follows easily from the
structure of the constraints where  $p_a$ 
can be eliminated as follows. First of all $D-1$ momenta
 $P_a^b p_b$ vanish strongly. On the other hand  $p_a\lambda^a$ can be solved using
 $\Omega$ and we obtain 
\begin{equation}
\lambda^a p_a=\pm \frac{1}{2}(
\lambda^b\lambda_b)\sqrt{h}\sqrt{1+h^{ij}\partial_i\phi\partial_j\phi} \ . 
 \end{equation}
   Further, $\tk^a,\psi$ vanish strongly so that unrestricted 
 variables are $\phi^a $ and conjugate veriables $\lambda_a$ while 
from $\Sigma$ we express $\lambda$ as a function of canonical variables.
Inserting these results into the Hamiltonian constraint $\mH_T$ we obtain the final result
\begin{eqnarray}\label{mHTM1}
\mH_T=\frac{2}{\sqrt{h}}\pi^{ij}\mG_{ijkl}\pi^{kl}-\sqrt{h}R
\pm p_\phi \sqrt{1+h^{ij}\partial_i\phi\partial_j\phi}+
\frac{1}{2}\sqrt{h}\lambda_a h^{ij}\partial_i\phi^a\partial_j\phi+
\frac{1}{2}\sqrt{h}V(\phi^a) \ . 
\nonumber \\
\end{eqnarray}
We derived remarkable result that shows that despite of the complexity of the extended model the Hamiltonian constraint still possesses linear dependence 
on the momentum $p_\phi$ that is  conjugate to the scalar field $\phi$
 as in original mimetic dark energy model.
Usually the presence of the linear momentum in the Hamiltonian is sign of an instability. However we can argue that this cannot be the case of the model
studied here. To see this in more details let us introduce auxiliary field $M$ with conjugate momentum $P_M$ that is the primary constraint and with the Poisson brackets 
\begin{equation}\label{pbM}
\pb{M(\bx),P_M(\by)}=\delta(\bx-\by) \ . 
\end{equation}
With the help of these fields we can rewrite the Hamiltonian constraint $\mH_T$
given above to the form 
\begin{equation}\label{mHTM}
\mH_T=\frac{2}{\sqrt{h}}\pi^{ij}\mG_{ijkl}\pi^{kl}-\sqrt{h}R
+\frac{p_\phi^2}{2\sqrt{h}M}+\frac{M}{2}\sqrt{h} (1+h^{ij}\partial_i\phi\partial_j\phi)+
\frac{1}{2}\sqrt{h}\lambda_a h^{ij}\partial_i\phi^a\partial_j\phi+
\frac{1}{2}\sqrt{h}V(\phi^a) \   
\end{equation}
which strongly resembles the Hamiltonian constraint for the dust that was
carefully analyzed in \cite{Brown:1994py}. In fact solving the equation of motion for $M$ which is equivalent to the requirement of the preservation of the primary constraint $P_M\approx 0$ we obtain 
$M^2=\frac{p_\phi^2}{h(1+h^{ij}\partial_i\phi\partial_j\phi)}$. Inserting this result into (\ref{mHTM}) we obtain (\ref{mHTM1}). The point is that the
scalar field part of the Hamiltonian constraint (\ref{mHTM}) is clearly positive definite on condition that $M>0$ and hence there is no sign of instability. For the case when $M<0$ we can certainly perform trivial canonical transformation $(M,P_M)\rightarrow (-M,-P_M)$ that preserves the Poisson 
bracket (\ref{pbM}) so that without lost of generality we can presume that $M>0$. 

On the other hand there is potentially another source of instability in this
theory which is the fact that the theory is linear in $\lambda_a$ and we argued above that this variable can be considered as the variable conjugate to $\phi^a$ due to the presence of the non-trivial Dirac brackets between them. We deal with this term in the same way as in the the case of the linear term in $p_\phi$. We introduce auxiliary field  $N_{ij}=N_{ji}$
 with inverse $N^{ij}$ and rewrite the term $\frac{1}{2}\sqrt{\lambda}
\lambda_a h^{ij}\partial_i\phi^a\partial_j\phi$ in the Hamiltonian constraint (\ref{mHTM})  as 
\begin{equation}
\frac{1}{2}\sqrt{h}\lambda_a h^{ij}\partial_i \phi^a \partial_j\phi
\rightarrow 
\frac{1}{4}\sqrt{h}\left(\lambda_a N^{ab}\lambda_b+N_{ab}
(h^{ij}\partial_i\phi\partial_j\phi^a)(h^{kl}\partial_k\phi
\partial_l\phi^b)\right) \  
\end{equation}
so that the Hamiltonian constraint 
(\ref{mHTM}) has the extended form 
\begin{eqnarray}\label{mHTMex}
\mH_T&=&\frac{2}{\sqrt{h}}\pi^{ij}\mG_{ijkl}\pi^{kl}-\sqrt{h}R
+\frac{p_\phi^2}{2\sqrt{h}M}+\frac{M}{2}\sqrt{h} (1+h^{ij}\partial_i\phi\partial_j\phi)+\nonumber \\
&+&
\frac{1}{4}\sqrt{h}
\left(\lambda_a N^{ab}\lambda_b+N_{ab}
(h^{ij}\partial_i\phi\partial_j\phi^a)(h^{kl}\partial_k\phi
\partial_l\phi^b)\right)
+ \frac{1}{2}\sqrt{h}V(\phi^a) \nonumber \\   
\end{eqnarray}

Next we introduce the constraints $P^{ab}\approx 0$, where  momenta $P^{ab}$  are conjugate to $N_{ab}$ 
with following Poisson brackets 
\begin{equation}
\pb{N_{ab}(\bx),P^{cd}(\by)}=\frac{1}{2}\left(
\delta_a^c\delta_b^d+\delta_a^d\delta_b^c\right)\delta(\bx-\by) \ . 
\end{equation}
Then the requirement of the preservation of the constraint $P^{ab}\approx 0$ implies the equation
\begin{equation}
-\lambda_c N^{ca}\lambda_d N^{db}+h^{ij}\partial_i\phi\partial_j\phi^a=0
\end{equation}
that can be solved as $\lambda_c N^{ca}=h^{ij}\partial_i\phi
\partial_j\phi^a$. Inserting this result into (\ref{mHTMex}) we reproduce the Hamiltonian constraint (\ref{mHTM}). Since we can demand that $N^{ab}$ is positive definitive exactly in the same way as in case of the variable $M$ we find that the scalar field contribution to the 
Hamiltonian constraint is positive definitive and hence there is no sign of instability which is certainly desired result. We also support this claim with the analysis of simpler model studied in the next section. 
\section{The Case of Single Scalar Field}\label{third}
In this section we focus on much simpler model when the number of 
additional scalar fields in the action (\ref{actHam}) is equal to one. In this 
 case it is convenient to  introduce a notation 
$\phi^1=\psi, \lambda_1=\omega$ so that 
\begin{eqnarray}
p_\phi=\lambda \sqrt{h}\nabla_n\phi+\frac{1}{2}\sqrt{h}
\omega \nabla_n\psi 
\ , \quad 
p_\psi=\frac{1}{2}\omega \nabla_n\phi \ , \quad   k_\lambda\approx 0 \ , \quad 
k_\omega \approx 0
\end{eqnarray}
and hence we easily  find an inverse transformation 
\begin{eqnarray}
\nabla_n\phi=\frac{2}{\sqrt{h}\omega}p_\psi \ , \quad 
\nabla_n\psi=\frac{2}{\sqrt{h}\omega}
\left(p_\phi-\frac{2\lambda}{\omega}p_\psi\right)
\nonumber \\
\end{eqnarray}
and corresponding Hamiltonian 
\begin{eqnarray}
H&=&\int d^3\bx (\pi^{ij}\partial_t h_{ij}+p_\phi \partial_t \phi+p_\psi\partial_t \psi-\mL)=
\nonumber \\
&=&\int d^3\bx (N\mH_T+N^i\mH_i) \ ,  \nonumber \\
\end{eqnarray}
where
\begin{eqnarray}
\mH_T&=&\frac{2}{\sqrt{h}}\pi^{ij}\mG_{ijkl}\pi^{kl}-\sqrt{h}R+
\frac{2}{\sqrt{h}\omega}p_\psi p_\phi-\frac{2\lambda}{
\sqrt{h}\omega^2}p_\psi^2
\nonumber \\
&+&\frac{1}{2}\lambda\sqrt{h}(1+h^{ij}\partial_i\phi
\partial_j\phi)+\frac{1}{2}
\sqrt{h}\omega h^{ij}\partial_i \psi\partial_j\phi+
\frac{1}{2}\sqrt{h}V(\psi) \ , \nonumber \\
\mH_i&=&-2h_{ik}D_j\pi^{kj}+p_\phi\partial_i\phi+
p_\psi\partial_i\psi \ .
\nonumber \\
\end{eqnarray}
Now requirement of the preservation of the momentum conjugate to $\lambda,\omega$ gives
\begin{eqnarray}
\partial_t k_\lambda&=&\pb{k_\lambda,H}=N\left(
\frac{2}{\sqrt{h}\omega^2}p^2_\psi-\frac{1}{2}\sqrt{h}
(1+h^{ij}\partial_i\phi\partial_j\phi)\right)=
N\Omega \approx  0 \ , \nonumber \\
\partial_t k_\omega&=&\pb{k_\omega,H}=
N\left(\frac{2}{\sqrt{h}\omega^2}p_\psi p_\phi
-\frac{4\lambda}{\sqrt{h}\omega^3}p^2_\psi-\frac{1}{2}
\sqrt{h}h^{ij}\partial_i\psi \partial_j\phi\right)
\equiv N\Sigma \approx 0  \  
\nonumber \\
\end{eqnarray}
while the requirement of the preservation of the momenta conjugate to 
$N,N^i$ again implies two secondary constraints $\mH_T\approx 0  \ , 
\mH_i\approx 0$.  We see that the constraint structure is much simpler than 
in case of general number of scalar fields $\phi^a$. In fact, it is very 
easy to determine the Poisson brackets between smeared form of the constraints
$\mH_T\approx 0 \ , \mH_i\approx 0$
\begin{eqnarray}
\pb{\bT_T(N),\bT_T(M)}&=&
\bT_S((N\partial_i M-M\partial_iN)h^{ij}) \ , \nonumber \\
\pb{\bT_S(N^i),\bT_S(M^i)}&=&\bT_S((N^i\partial_iM^j-M^i\partial_iN^j)) \ , 
\nonumber \\
\pb{\bT_S(N^i),\bT_T(M)}&=&\bT_T(N^i\partial_iM) \  . \nonumber \\
\end{eqnarray}
Of course, we still have to ensure that $\mH_T$ is the first class constraint. This can be easily done using the prescription presented in previous section so that we will not repeat it here but we simply sat that 
$\mH_i\approx 0 \ , \tmH_T\approx 0$ are first class constraints which 
is a reflection of the  full diffeomorphism invariance of the theory. 

The number of physical degrees of freedom is obtained via Dirac's
formula: (number of canonical variables)/2 $-$ (number of first class
constraints) $-$ (number of second class constraints)/2. Using 
Hamiltonian and spatial diffeomoprhism constraints we can eliminate 
eight number of degrees of freedom from the gravitational sector 
with twelfth variables $h_{ij},\pi^{ij}$ so that we obtain four phase
space degrees of freedom corresponding to massless graviton. Further, 
$k_\lambda,k_\omega$ vanish strongly since they are the second class constraints
with $\Omega $ and $\Sigma$ that   can be solved for $\omega$
and $\lambda$. Explicitly, from $\Omega$ we find
\begin{equation}
\omega =
\pm\frac{2}{\sqrt{h}
	\sqrt{(1+h^{ij}\partial_i\phi\partial_j\phi)}}p_\psi \ ,
\end{equation}
while from  $\Sigma=0$ we express $\lambda$ as 
\begin{equation}
\lambda=\frac{\sqrt{h}\omega^3}{4 p_\psi^2}
\left(\frac{2}{\sqrt{h}
\omega^2}p_\psi p_\phi-\frac{1}{2}\sqrt{h}h^{ij}
\partial_i\psi\partial_j\phi\right) \ . 
\end{equation}
Note that the dependence on $\lambda$ disappears from the Hamiltonian 
since the Hamiltonian constraint is linear in $\lambda$. 
Inserting these results into the Hamiltonian constraint we obtain 
\begin{eqnarray}\label{mHsingle}
\mH_T=\frac{2}{\sqrt{h}}\pi^{ij}\mG_{ijkl}\pi^{kl}-\sqrt{h}R \pm
\sqrt{1+h^{ij}\partial_i\phi\partial_j\phi}p_\phi
\pm \frac{1}{\sqrt{1+h^{ij}\partial_i\phi
\partial_j\phi}}p_\psi h^{ij}\partial_i\psi \partial_j\phi+
\frac{1}{2}\sqrt{h}V(\psi) \  \nonumber \\
\end{eqnarray}
and we see that the Hamiltonian constraint is linear in $p_\phi$ and $p_\psi$. However introducing two auxiliary fields $M$ and $K$ with conjugate momenta $P_M\approx 0 \ , P_K\approx 0$ we can rewrite this Hamiltonian constraint into the form 
\footnote{This is possible on condition when $\partial_i\psi\neq 0 , \partial_i\phi\neq 0$. Clearly when either $\phi$ or $\psi$ depend on time only term linear in $p_\psi$ is zero and potential problem with instability disappears.}
\begin{eqnarray}\label{mHsingleex}
\mH_T&=&\frac{2}{\sqrt{h}}\pi^{ij}\mG_{ijkl}\pi^{kl}-\sqrt{h}R +
\frac{1}{2\sqrt{h}M}p_\phi^2+\frac{1}{2}\sqrt{h}M(1+h^{ij}
\partial_i\phi\partial_j\phi)+\nonumber \\
&+& \frac{1}{2\sqrt{h}K}p_\psi^2+
\frac{1}{2}\sqrt{h}K\frac{(h^{ij}\partial_i\phi\partial_j\psi)^2}
{1+h^{ij}\partial_i\phi\partial_j\phi}+
\frac{1}{2}\sqrt{h}V(\psi) \ .  \nonumber \\
\end{eqnarray}
Repeating the same arguments as in previous section we can presume that
$M>0,K>0$ without lost of generality so that contribution
from the scalar field in  (\ref{mHsingleex})
is positive definite  and hence it is bounded from below which is satisfactory fact.

Finally we determine schematic form of Dirac brackets between canonical variables. 
Let us denote the second class constraints as $\Psi_A=(k_\lambda,k_\omega,
\Omega,\Sigma)$.  Then it ie easy to see that the matrix of the Poisson brackets between second class constraints has the form 
\begin{equation}
\Omega_{AB}=\left(\begin{array}{cc}
0 & A \\ 
-A & B \\ \end{array}\right)
\end{equation}
with inverse 
\begin{equation}\label{Omegainv}
\Omega^{-1}=\left(\begin{array}{cc}
A^{-1} B A^{-1} & -A^{-1} \\
A^{-1} & 0 \\ \end{array}\right) \ , 
\end{equation}
where $A,B$ are $2\times 2$ matrices. Then the Poisson brackets between canonical variables and second class  constraints have schematic form
\begin{eqnarray}\label{canPsi}
\pb{h_{ij},\Psi^T}=(0,0,0,0) \ , \quad
\pb{\pi^{ij},\Psi^T}=(0,0,*,*) \ , \nonumber \\
\pb{\phi,\Psi^T}=(0,0,0,*) \ , \quad 
\pb{p_\phi,\Psi^T}=(0,0,*,*) \ , \nonumber \\
\pb{\psi,\Psi^T}=(0,0,*,*) \ , \quad 
\pb{p_\psi,\Psi^T}=(0,0,0,*) \ . \nonumber \\
\end{eqnarray}
Then we can easily calculate Dirac bracket for canonical variables. 
Let us demonstrate this calculation on following examples 
\begin{eqnarray}
\pb{h_{ij},\pi^{kl}}_D&=&\pb{h_{ij},\pi^{kl}}-
	\pb{h_{ij},\Psi^T}\Omega^{-1}\pb{\Psi,\pi^{kl}}=\pb{h_{ij},\pi^{kl}} \ ,
	\nonumber \\
\pb{\pi^{ij},\pi^{kl}}_D&=&-\pb{\pi^{ij},\Psi^T}\Omega^{-1}\pb{\Psi,\pi^{kl}}=0  \ \nonumber \\	
	\end{eqnarray}
as follows from (\ref{Omegainv}) and (\ref{canPsi}). In the same way we find that all Dirac brackets coincide with corresponding Poisson brackets between canonical variables.

\section{$f(\Box\phi)$ Model}\label{fourth}
In this section  we consider Hamiltonian formulation of the model that was proposed
 in \cite{Chamseddine:2016uef,Chamseddine:2016ktu}. This model is defined by the action 
 \footnote{The special case of this model with $f(\Box\phi)=\gamma
 (\Box\phi)^2$ was studied recently in \cite{Ramazanov:2016xhp}
where it was introduced as the covariant action for the IR limit of the projectable Ho\v{r}ava-Lifshitz gravity.}
\begin{equation}\label{Sgff}
S=\frac{1}{2}\int d^4x\sqrt{-g}
\left[ R(g_{\mu\nu}) -\lambda\left( 1+g^{\mu\nu}
\nabla_\mu\phi\nabla_\nu\phi \right)+f(\Box \phi) \right] \ ,
\end{equation}
where $f$ is given in (\ref{fBox})  and $\Box=\frac{1}{\sqrt{-g}}
\partial_\mu[\sqrt{-g}g^{\mu\nu}\partial_\nu]$. In order to proceed
to the Hamiltonian formalism we introduce two auxiliary fields and rewrite
the action into the form 
\begin{eqnarray}\label{Sgff2}
S&=&\frac{1}{2}\int d^4x\sqrt{-g}
\left[ R(g_{\mu\nu}) -\lambda\left( 1+g^{\mu\nu}
\nabla_\mu\phi\nabla_\nu\phi \right)+f(A)+B(A-\Box\phi) \right]=\nonumber \\
&=&\frac{1}{2}\int d^4x\sqrt{-g}
\left[ R(g_{\mu\nu}) -\lambda\left( 1+g^{\mu\nu}
\nabla_\mu\phi\nabla_\nu\phi \right)+f(A)+BA
+\sqrt{-g}\partial_\mu B g^{\mu\nu}\partial_\nu\phi \right]=\nonumber \\
&=&\frac{1}{2}\int d^4x\sqrt{-g}
\left[ R(g_{\mu\nu}) -\lambda\left( 1+g^{\mu\nu}
\nabla_\mu\phi\nabla_\nu\phi \right)-U(B)
+\sqrt{-g}\partial_\mu B g^{\mu\nu}\partial_\nu\phi \right] \ ,\nonumber \\
\end{eqnarray}
where in the last step we solved the equation of motion for $A$ that has
the form 
\begin{equation}
\frac{df(A)}{dA}+B=0 
\end{equation}
and we presumed solution in the form $A=\Psi(B) \ , f' (\Psi(B))=-B$. Inserting this result back to the action we obtain the last form of the action with 
the potential $U(B)$ defined as 
\begin{equation}
U(B)=-f(\Psi(B))-B\Psi(B) \ . 
\end{equation}
Note the crucial difference between the action (\ref{Sgff2}) and the action studied in previous section 
which is an absence of the Lagrange multiplier  $\omega$. This fact will be very important for the structure of the constraints as we will see below. In order to proceed to the
Hamiltonian analysis of this action we write it in $3+1$ formalism
\begin{eqnarray}\label{Sfmodel}
S&=&\frac{1}{2}\int dt d^ 3\bx N\sqrt{h}
[K_{ij}\mG^{ijkl}K_{kl}+R+\lambda(-1+\nabla_n\phi\nabla_n\phi
-h^{ij}\partial_i\phi\partial_j\phi)-
\nonumber \\
&-&\nabla_n B\nabla_n\phi+h^{ij}\partial_i B\partial_j\phi-U(B)]
\ . 
\nonumber \\
\end{eqnarray}
From (\ref{Sfmodel}) we easily find
\begin{eqnarray}
\pi^{ij}=\frac{1}{2}\sqrt{h}\mG^{ijkl}K_{kl} \ , \quad  
p_B=-\frac{1}{2}\sqrt{h}\nabla_n\phi \ , \quad 
p_\phi=\sqrt{h}\lambda\nabla_n \phi-\frac{1}{2}\sqrt{h}\nabla_n B
\nonumber \\
\end{eqnarray}
so that the Hamiltonian has the form 
\begin{eqnarray}
H&=&\int d^3\bx (\pi^{ij}\partial_t h_{ij}+
p_\phi \partial_t\phi+p_B\partial_t B-\mL)=
\int d^ 3\bx (N\mH_T+N^i\mH_i) \ , \nonumber \\
\end{eqnarray}
where 
\begin{eqnarray}
\mH_T&=&\frac{2}{\sqrt{h}}\pi^{ij}\mG_{ijkl}\pi^{kl}-\frac{1}{2}\sqrt{h}R
+\frac{1}{2}\sqrt{h}\lambda(1+h^{ij}\partial_i\phi\partial_j\phi)
\nonumber  \\
&-&\frac{2}
{\sqrt{h}}p_\phi p_B-\frac{2\lambda}{\sqrt{h}}p_B^2-
\frac{1}{2}\sqrt{h}h^{ij}\partial_i B\partial_j\phi+\frac{1}{2}\sqrt{h}U(B) \ , 
\nonumber \\
\mH_i&=&-2h_{ik}D_j\pi^{jk}+p_\phi\partial_i\phi+
p_B\partial_i B \ . \nonumber \\
\end{eqnarray}
Now the preservation of the primary constraints $\pi_N\approx 0 , \pi_i\approx 0$
again implies two secondary constraints $\mH_T\approx 0 \ , \mH_i\approx 0$ 
while the requirement of the preservation of the constraint $p_\lambda\approx 0$
implies
\begin{equation}
\partial_t p_\lambda=\pb{p_\lambda,H}=
N\left(-\frac{1}{2}\sqrt{h}(1+h^{ij}\partial_i\phi\partial_j\phi)+
\frac{2}{\sqrt{h}}p_B^2\right)\equiv N\Sigma \approx 0 \ .
\end{equation}
Now we have to require that the constraint $\Sigma(\bx)$ is preserved during the time evolution of the system. To do this we have to calculate
\begin{eqnarray}
\partial_t \Sigma&=&\pb{\Sigma,\int d^3\by N\mH_T(\by)}=
N\left( 2\sqrt{h}\partial_i \phi h^{ij}\partial_j(\frac{p_B}{\sqrt{h}})
-2\frac{p_B}{\sqrt{h}}\partial_i(h^{ij}\partial_j\phi) \right.\nonumber \\
&+&\left.\frac{1}{2}
h_{ij}\pi^{ij}(1+h^{kl}\partial_k\phi\partial_l\phi)+\frac{2}{h}
h_{ij}\pi^{ij}p_B^2+2\partial_i\phi\partial_j\phi \pi^{ij}-
\partial_i\phi\partial_j\phi h^{ij}h_{kl}\pi^{kl}-2p_B\frac{dU}{dB}\right)
\nonumber \\
&\approx &  
N\left( 2\sqrt{h}\partial_i \phi h^{ij}\partial_j(\frac{p_B}{\sqrt{h}})
-2\frac{p_B}{\sqrt{h}}\partial_i(\sqrt{h}h^{ij}\partial_j\phi)+2\partial_i\phi\partial_j\phi \pi^{ij}+h_{ij}\pi^{ij}-2p_B\frac{dU}{dB }\right)=
N\Sigma^{II} \ ,
\nonumber \\
\end{eqnarray}
where in the last step we used constraint $\Sigma $. We see that  in order to preserve constraint $\Sigma$ during the time
evolution of the system we have to require that either $N$ or $\Sigma^{II}$
vanish. Clearly the first condition is too strong and it implies 
singular metric so that it is natural to demand an existence of the
new constraint $\Sigma^{II}\approx 0$. 
In other words we have two second class constraints $\Sigma(\bx)\approx 0 \ , 
\Sigma^{II}(\bx)\approx 0$.Considering $\mH_T,\mH_i$ we extend them in the same way as in previous sections and we find that $\tmH_T,\tmH_i$ are first class constraints. 

Let us now solve the second class constraints $\Sigma,\Sigma^{II}$.
The  first one can be solved for $p_B$ while the second one can be solved for $B$ if we presume an existence 
of the inverse function to $\frac{dU}{dB}$. Then we can write $B=B(h_{ij},\pi^{ij},p_\phi,\phi)$ and hence the Hamiltonian constraint
$\mH_T$, after solving the second class constraints, has the form 
\begin{eqnarray}
\mH_T&=&\frac{2}{\sqrt{h}}\pi^{ij}\mG_{ijkl}\pi^{kl}-\frac{1}{2}\sqrt{h}R
\pm p_\phi\sqrt{1+h^{ij}\partial_i\phi\partial_j\phi}-\nonumber \\
&-&\frac{1}{2}\sqrt{h}h^{ij}\partial_i B(h_{ij},\pi^{ij},\phi,p_\phi)\partial_j\phi+
\frac{1}{2}\sqrt{h}U(B(h_{ij},\pi^{ij},\phi,p_\phi)) \ 
\nonumber \\
\end{eqnarray}
so that we again find that the Hamiltonian constraint is linear in 
$p_\phi$ and we can make it bounded from bellow exactly as in the previous sections.  On the other hand we see that  there are no additional 
degrees of freedom as we could expected from the presence of 
the d'Alembertian $\Box$
in the action. This result can be considered as an confirmation of the
claim presented in   
\cite{Chamseddine:2016uef}. Naively we could  expect  that due to the presence of the higher derivative operator in the action 
(\ref{Sgff}) Ostrogratsky instability occurs. On the other hand this is strictly true in theory which is non-degenerate while the action 
(\ref{Sgff}) is degenerate theory which leads to the presence of the constraints in the Hamiltonian formalism that eliminate additional degrees of freedom as we showed above. 

Now we briefly mention the form of the Dirac brackets between 
canonical variables. From the form of second class constraints 
$\Psi_A=(\Sigma,\Sigma^{II})$ we easily find that it has the form
\begin{equation}
\Omega=\pb{\Psi_A,\Psi_B}=\left(\begin{array}{cc}
0 & A \\
-A & B \\ \end{array}\right) \ 
\end{equation}
so that inverse matrix has the schematic form 
\begin{equation}
\Omega^{-1}=\left(\begin{array}{cc}
A^{-1} B A^{-1} &  -A^{-1} \\
A^{-1} & 0 \\ \end{array}
\right) \ .
\end{equation}
We again introduce  the notation 
\begin{eqnarray}
\pb{h_{ij},\Psi^T}&=&(0,*) \ , \quad 
\pb{\pi^{ij},\Psi^T}=(*,*) \ , \nonumber \\
\pb{\phi,\Psi^T}&=&(0,0) \ , \quad \pb{p_\phi,\Psi^T}=(*,*) \ . 
\nonumber \\
\end{eqnarray}
Now since $\pi^{ij}$ and $p_\phi$ have non-zero Poisson brackets
with the primary constraint $\Sigma$ we obtain that the structure
of Dirac brackets is more complicated. For example 
\begin{eqnarray}
\pb{h_{ij},\pi^{kl}}_D=\pb{h_{ij},\pi^{kl}}+\pb{h_{ij},\Sigma^{II}}
A^{-1}\pb{\Sigma,\pi^{kl}} \ . \nonumber \\
\end{eqnarray}
In the same way we can show that there are non-zero Dirac brackets
$\pb{\pi^{ij},\pi^{kl}}_D,\pb{p_\phi,\pi^{ij}}_D$ and so on. In other words
the phase space has very complicated structure as opposite to the original
form of the mimetic theory. 
\section{Conclusions}\label{fifth}
We have studied inhomogeneous mimetic model proposed in 
\cite{Chamseddine:2016uyr} from Hamiltonian point of view. We argued that
in case of general number of additional scalar fields there are 
new primary constraints that makes the analysis rather complicated. 
On the other hand we have shown that despite of this fact the 
Hamiltonian constraint is linear in momentum $p_\phi$ conjugate to 
scalar field $\phi$ which signals possible instability of this model 
which is the same situation as in case of the original mimetic model. 
However we also argue that it is possible to rewrite the scalar part of the Hamiltonian constraint to have the same form as in case of the dust which is well defined system \cite{Brown:1994py} since the Hamiltonian 
constraint is quadratic and bounded from below. 

In the next part of this paper we performed canonical analysis 
of the model proposed in  \cite{Chamseddine:2016uef}. We 
determined structure of the constraints  and we again showed that the Hamiltonian constraint is linear in the momentum $p_\phi$ after solving
second class constraints. We also argued that the Dirac brackets on the reduced phase space have non-trivial structure which makes further analysis of this theory rather complicated. On the other hand we mean that it would be very interesting to analyze mimetic theory and its modification following
seminal papers  \cite{Thiemann:2006up,Brown:1994py}. We return to this problem in future. 

\acknowledgments{This  work  was
supported by the Grant Agency of the Czech Republic under the grant
P201/12/G028. }

\end{document}